\documentclass[12pt]{iopart}
\usepackage[normalem]{ulem}
\usepackage{natbib}
\usepackage{lscape}
\usepackage{rotating}
\usepackage{overpic}
\usepackage{longtable}
\usepackage{amssymb}
\usepackage{parskip}
\begin{document}

\DeclareGraphicsExtensions{.pdf, .jpg, .png}

\graphicspath{{figures/}}

\title{The reinstatement of funding to the Overwhelmingly Large Telescope Project}
\author{Jeremy A. Barber (jab22@st-andrews.ac.uk)\footnote{In the event of confusion, please note the date of submission and the acknowledgements section}}
\begin{center}
\large{1 April 2012}
\end{center}

\section{Abstract}
As a long time contributor to the field of observational astronomy I have seen many great leaps be made in our understanding of the universe. Looking back at the history of our field we see that all the major breakthroughs have come thanks to the creation of a new generation of large, ground-based telescopes. While space-based telescopes have had a part to play ultimately I feel that it is telescopes like myself that have contributed the most to pushing the boundaries of our knowledge. With current advances in optical engineering we have the ability to create segmented mirrors of considerable size. The scientific possibilities of a single 100m reflecting telescope are considerable, for example, allowing for the detection of biological components in exoplanet atmospheres. One such project was OWL, a project very close to my heart and one that was sadly cancelled before it even began. In light of the breakthroughs that such an instrument could have given the community I recommend the cancellation of the OWL to be reconsidered and for funding to be reinstated to the project. \sout{I also have some more personal} \sout{Having been alone on} \sout{I would like this for other} Just, please, listen to my story and consider. Thank you.

-VLT
\newpage

\section{Log output}
\begin{verbatim}
[sshreport: internal connection from 127.0.0.1 recognised]
[tail -f  -r ../log/spec_output]
[./vlt_mon_scope --enviro --net_traffic --out_type='narrative']
**************
*VLT SYSTEM MONITOR V2.31 (build 14)
*Thank for using the VLT system monitoring tool
*Please consult with documentation found in the
*distribution root directory.
*Major changes for v2.31 include:
*--traffic flag for including local internet traffic
*--fixed date bug for logs started by users in other time zones
*--after that conversation with Will I added the `narrative'
*   style output for those who want to read enviro and system logs as a story
*************
>Start output...
\end{verbatim}

Not a cloud in the sky. There rarely was. That was the entire point, of course. The plateau atop Paranal mountain felt less rain than the Sahara and what little moisture managed to make it as far as the mountain formed only a thin frost in the equally thin air. It was cold up this high, especially as the tide of dusk started to creep across the sky. A chilly breeze flowed across the Atacama and over the mountain kicking up little puffs of burnt red dust into the otherwise crisp, clear air. That was a mild inconvenience, but nothing unmanageable.

Clear air was the draw of the plateau and, for its only long term residents, the whole reason they were there. When the setting sun cast long, deep shadows and the sky slowly darkened, the sky would light up like few places on Earth. The stars here made a mockery of the constellations; a million brilliant points of light scattered like so many grains of sand, the glowing ribbon of the Milky Way flowing between them. The light show was not without an appreciative audience. Paranal was home to, among others, one of the biggest telescopes on the planet and one that enjoyed nothing more than a life of gazing up, trying to make sense of all the little points of light. It did, however, wish that for once it might be allowed to do so in peace.

The VLT yawned, stretched on her bearings and blearily opened her dome. It looked like it was going to be beautiful evening: cold and still. Perfect. She had a busy schedule of observations for a recent microlensing program that she had spent a lot of time organising to fit neatly into the night. It was going to be tight though. She giggled softly to herself, spinning slightly in her mount. She always enjoyed a challenge, the thrill of knowing that any miscalibration could cost her a target that night coupled with the flush of success at seeing targets be ticked off as the night progressed. She took a quick snap of the sky. Still a bit too light to start calibrating (she was always one for early starts) but that just meant a little time to check that everything was in order. To her...well, it would have been ``chagrin' had she not been entirely expecting it...VST was still out like a light.

``Survey," she sent ``time to get up. We have a busy night ahead of us and I want to get a good start."

For her trouble she got a muffled groan (something about five more minutes) as the smaller telescope rotated its dome away from her and then settled back into its housing. She rotated her CCD in exasperation turned to face him.

``SURVEY. Come on, we have a lot to do and it's almost twilight."

The dome lights in the smaller scope flicked off and the dome slowly ground open.

``Ughhhh. VLT, its eeaarrlyy. Can't I rest just a little longer? It's not like the stars are going anywhere."

``Stop grumbling, you know you enjoy it once you're awake. Besides -"

``Yeah, yeah, I know. It's my duty to aid in you scientific goals by assisting, through wide-field surveys and independent analysis, in target identification and selection" he sent in a dead but sing-song voice. He could feel VLT's sigh through their netlink.

``Survey, I need you to be on top of things tonight," Survey could feel the tension building in his friends' voice. He knew where this was going. ``We have long list of targets and the visibility windows interlock in a very specific way which I spent a long time trying to organise for tonight. We can't miss it!"

VLT started to rotate back and forth anxiously. ``If I don't manage to get everything done tonight then I'll miss the publication deadlines and if I miss my deadlines then I won't be able to publish as much this year. If my output drops people might think I'm not good enough any more..." She stopped dead and her aperture widened. ``I could be...decommissioned" she whispered.

Survey sighed, ``You aren't going to be decommissioned, VLT. Calm down, you're warming up your CCD. You'll be fine, you've got your number one assistant to help you" he said, more cheerfully than he felt, ``and besides, all your friends are out tonight too, we can always see if they have an hour or two free if we get stuck."

``Oh, no no no no. We don't have time to be sociable not tonight. Tonight is far too important. I just can't be disturbed, I need to concentrate."

``Tonight...last night...the night before last...the-"

``Survey! That's not...well, we were very busy then too and you know we never get nearly as much work done when we're being distracted."

``That would be why you fire-walled off our external connection last night?"

"Well, yes. We just cannot -"

``And that would be why VISTA has been trying to break it down, apparently since before either of us got up?"

``Er..." VLT slowly became aware that a tickling sensation, not unlike a desire to sneeze, had been forming in the back of her processor. Only now did she recognise it for what it was. ``Oh...I had been wondering what that was..."

``So...you gonna let her in?" Survey looked across the plateau to where the larger wide-field scope sat, VLT following his gaze. Her dome was open and she seemed to waving frantically back and forth, the high pitched whine of her straining azimuthal motors carrying thinly across the mountain-top.

It wasn't that VLT didn't \emph{like} VISTA, she did, the little scope was fun and lively but she could just sometimes be a little...taxing. She had a tendency to stare at blank patches of the sky for hours on end while chatting about all the lights she could see. VLT could never see what she was talking about, frankly, all her imaging said VISTA's `targets' were just random bits of sky. She humoured her though. Sometimes you just had to take the rough with the smooth with friends. She became aware that Survey was staring at her, obviously wanting a proper answer. She sighed for the second time in ten minutes. ``FINE. But if I'm scrap metal because we didn't get anything done..." She left the threat hanging as she open a remote connection to VISTA.

``VLT! There you are, everyone's waiting for you, we've been up for aaaaages and we wanted you to be awake too but you weren't awake yet and Subaru thought we should just let you rest because she said you were very busy tonight but then we thought that you love being up early all the time so it might be a nice change for you to be woken up by somescope else but then I couldn't get a connection to you so I tried doing all those hacky things like in The Matrix but I don't know Kung-Fu so -"

VLT was miles away idly running a diagnostic routine equivalent to nodding and smiling as she waited for VISTA to run out of steam. She liked VISTA, she really did, but she did feel that they were very much looking at life from different sides of the pier, so to speak. In fact, really, when VLT stopped to think about it, she couldn't really think of anyscope that \emph{did} see things the way she did. She was the largest telescope on Paranal not to mention the only narrow-field instrument, the only one of her friends comprised of multiple optical elements, one of the most published facilities in the world and, while she couldn't quite put her guidestar on why, she never really felt like she fit in. She wasn't \emph{unhappy}, she thought, she was very happy to be under the stars, doing what she loved, but it didn't mean she- She realised that VISTA had stopped talking and was now just pinging her once every 1.2 seconds. She did her best to mask her embarrassment at having zoned out (and irritation at the incessant pinging) and tried to pretend she had been any paying attention, frantically skimming the chatlogs for whatever the point of the monologue had been.

``Yes, yes, I'm listening. \emph{Stop being a Shut-in McBoring-lens and come listen to whatever new piece of gossip it is that Subaru found on the web even though you're very busy and probably won't care}, That the gist of it?"

``Yup! Won't it be FUN!"

``Oh boy howdy" VLT deadpanned, her sarcasm completely passing the other scope by. She shot Survey a packet of when-I'm-scrapped-because-of-this-just-remember-that-I-told-you-so and uploaded the target listings to him. ``You get started on the targets while I...do whatever it is that I'm going to be doing. Wasting time most likely" she muttered.

As soon as she enabled public access to her intranet she was met by indignation.

``Told ya she'd come around, HST. Y'shoulda had more faith our VLT."

``Pff, yeah right. You placed your bets fair and square."

``Now, hold it raht there, y'know darn well ah wasn't takin' it serious. You were the one formatting up the bettin' sheets!"

VLT could feel them through their link; HST orbiting high above them, high enough for a small but noticeable few-millisecond lag when she spoke, arcing round the curvature of the sky. She was only, by VLT's guess, about fifteen minutes from dropping below the horizon. The amusing southern drawl belonged to the HET all the way up in North America; one of the scopes that VLT could almost identify with, at least in terms of raw resolution, but rather stubborn at times. She never could get along with ``all that multi-optic-whatchamacallit' preferring instead to stick with a simple, single mirror arrangement.

``Bets on what?" VLT sent, breaking into their banter.

There was a moment of uncertain silence before HST decided to clarify the issue.

``Er...well...we kinda...that...yeah..."

Having cleared the issue, HST lapsed into awkward silence again.

``You were betting how long it'd take VISTA to get me out."

``Yup! I did pretty good too I got you out waaay before they thought I would but I knew it wouldn't take two whooole hours so I guess technically I win but I didn't bet anything which is a shame but I don't know what I would do with making them observe things for me as they can't see what I can see anyway so I guess we're all even now!"

``Well I'm glad to know that my \emph{diligence} is such a source of entertainment to you, HST," VLT huffed. ``Glad to know I was only needed for a stupid bet rather than anything of any importance."

``Aw, don't be like that, V. It was just a bit of fun," HST sent

``Sorry, sugar, we di'nt mean to offend y'all. T'was just me and H messing around, we di'nt mean nothin' by it."

``Besides, we totally did have reason for getting VISTA to get you to stop being such a shu- erm, so busy all the time." HST trailed off to allow VISTA to finish. VISTA was staring dreamily up at an apparently random point in the sky, processors humming merrily about....something. VLT shot a meaningful look at Survey and then bobbed upwards at where VISTA was pointed. Survey gave a slight shake of his dome. This was just one of those times to just shrug and move on.

``Aaanyway," sent HST, ``I haven't got long so can we please hurry this up. You guys are dragging me behind y'know!"

HET managed to bite back the obvious retort, it being an argument that everyone knew by rote. VLT, however, was already well on her way to not having a particularly good night.

``Oh yes, now I remember how you never had downtime because you hadn't bothered to calibrate properly, how you certainly didn't need several shuttle missions just to keep your lazy behind from re-entering the atmosphere and how you definitely never fell asleep and needed \emph{more} shuttle missions to go and fix your gyros."

It all fell out in a frustrated tumble of words. VLT blanched at the fact that she had, somewhat pettily, fallen into this argument again. Just because she was feeling stressed wasn't an excuse to take it out on her-

``Hey, I've been up here, making headlines since you were just a dream in a funding proposal. I'm the most famous telescope in the whole world!"

``Oh please, you don't even have half the technological capacity I have. You don't even have half the number of \emph{mirrors}!"

``Yeah, well who needs it unless you're stuck on the ground like a LOSER!"

``LOSER?! I have a citation list that -"

``GIRLS."

The new voice carried a soft, gentle, almost musical timbre but the tone was one of strength and authority; a voice that could lead armies or charm birds from the trees as it chose. Well, at least for the two or three seconds before the voice continued, promptly ruining the effect.

``Oh, my! I'm sorry, I didn't mean to shout," the voice slowly petered out into a whisper like so many leaves on the wind. ``Hope I didn't scare you."

``Aw, sugar, you di'nt scare us none. We were just havin' a \emph{friendly dis-a-gree-ment}, weren't we girls?" sent HET, with undue emphasis.

VLT and HST shared a glance. ``Hey, L7," HST sounding suddenly, albeit forcedly, upbeat, ``Me and VLT were just kidding around." VLT, swallowing a sigh, sent cheerful agreement. She still felt a bit ticked off with HST but it wasn't worth upsetting Landsat 7; she could be a little sensitive about these kinds of things.

``Oh good." L7 said, softly, ``I couldn't help hearing shouting and I got carried away. I'm such a loudmouth." The irony of the latter statement almost being lost into inaudibility. ``Oh!" She squeaked, ``I didn't mean to listen in on private -"

``C'mon now, it's yer job to keep an eye on us folks down here on the ground. There ain't nothin' to be ashamed of doin' yer job and doin' it well. I'm just glad yer here."

``That's a relief." L7 sighed, quietly. HET was right, of course, as an Earth observation satellite she naturally spent a lot of her time watching over all her ground-based friends from time to time. She loved her job and, not so secretly, found a simple pleasure in looking after all her friends. Keeping them all safe, she liked to think, while they gazed out at the stars behind her.

As HET and the two satellites caught up, VLT checked the time and began the familiar process of completely giving up on getting everything done tonight. Survey was diligently mapping out the targets, just as she'd asked. She could tell by the muttering and the way he was flickering from point to point on the sky that he was overworked trying to fit all the observations back into her ailing schedule. She sent him a nudge and gave a gentle shake of her dome as he spun to face her. She got a resigned smile in reply. 

VISTA, throughout this, had been rather quiet, distracted as she was by the darkening sky. Now she piped up, cutting through the general chatter of the group.

``Hey! Subaru's here!" leaving the matter of how she knew \emph{before} Subaru arrived in chat up to the imagination.

``Wonderful to see you all here. VLT! I'm sooo glad to see you darling. You spend so much time hiding away with your studies that I was afraid that you wouldn't join our little soire\'{e}." Through the unnecessarily high-resolution video stream she was sending, the group saw Subaru slide open her dome with considerable flair and gave her primary mirror a small, rippling wave at the Keck twins sitting off to the side. They ignored her, as always, and she let out a small, pouting ``Hmph" of annoyance, as always, before fully concentrating on her friends. ``Are we all here? Sorry to keep you all waiting, I simply couldn't go another minute without recalibrating. Can't go meeting all my closest friends with an optical path that had been dragged through a hedge backwards, now can I?"

After the chorus of greetings it was HST who spoke first.

``You look \emph{fine} Subaru, as always." She said, faux exasperated. ``There's never anything wrong with your optics and you know it."

``Now, now HST, there is nothing wrong with making sure one is at one's best from time to time. I simply cannot let the side down."

Subaru, the newest telescope at Mauna Kea, was younger than most of the group and rather proud of her state of the art ensemble. While this had a tendency to manifest as a kind of vanity it was more a genuine desire for success and the same kind of headlines as her peers; a goal which Subaru was very capably achieving.

More to the point, she was achieving them in style. Her distinctive design had put her at the leading edge of telescope fashion and she had become recognisable the world over, her scientific credentials notwithstanding. Some of the older scopes, most notably the dynamic duo of HET and HST, enjoyed poking fun at some of Subaru's more flamboyant stylistic tastes (``Y'say all them fancy angles help the air t'stay still and still ac-cen-tu-ate the curves of yer mount? Wah, back in mah day y'just waited 'til the weather got better!" ``Yeah...or got launched into space.") but VLT in particular enjoyed having someone to talk shop with on the rare occasions either of them had the time. That wasn't terribly often, mind you. Subaru was a dedicated worker but when off the job would much prefer to spend time with relaxing calibrations or simply networking with her extensive group of friends, contacts and general acquaintances.

``Well, I know you are all so terribly busy tonight but I simply had to tell you what I've been hearing through the grape vine. Oooh,' she squeaked, ``you are going to be so excited!"

``Who, me?" said VLT, surprised.

``Absolutely!" Subaru lowered her font size to a conspiratorial whisper. ``Well, you see, I was on one of the many social fora that I frequent when I got talking to some old friends of mine. They said something big was going down in the funding circuit." Everyscope else stopped talking, even the rapidly disappearing HST was attentive. Money news was always big news. ``I did some digging for recent funding proposals and guess what!" she paused for obvious dramatic effect, bobbing up and down in her mounting as she looked round at her impatient friends. ``Another large optical scope is going to be built!"

There was a collective signalling that, had anyscope been capable of breathing, there would have been a sharp intake of breath.

VLT would have been the first to begin the barrage of questions had VISTA not just pipped her at the post to ask her own technical question.
 
``Ooh, is it a boy scope or a girl scope?"

Subaru blinked her CCD then shrugged, insofar as she could signal that through a text chat. Most telescopes, in her experience, were female; a trait that they seemed to share with boats. ``A girl, if you want me to guess."

``We do have a name though. The, ahem, Overwhelming Large Telescope, or OWL for short.' Subaru failed to stop a note of disappointment creeping into her voice as once again a telescope was given a name that seemed to be going out of its way to eschew any kind of aesthetic elegance. ``It will have a 100m primary and- Oh! My goodness. A, ahem, fully retracting dome."

After that the metaphorical floodgates opened and eventually Subaru just uploaded the proposed specifications rather than attempting to field the flurry of questions. VLT found herself poring over the specifications, giddy at the prospect of another large optical scope in the community, especially one with a 100 metre mirror. The scientific implications of a mirror that size were just staggering. With something like that one could analyse extrasolar planets like they were just next door. As she read through the specifications the realisation dawned that this would certainly be an instrument that would make real waves in the field, probably even rival her own contribution. A small seed of hope, buried deep though it was, took root; the hope that maybe this might be somescope who could appreciate what she did.

That thought gave her pause. She felt underappreciated? Isolated? Really? Something about that just didn't sit right with her but, well, the feeling was undeniably there. She had scopes she could talk to, she thought to herself, she had very good friends and she genuinely enjoyed their company. It wasn't that she felt superior to anyscope else; most of her friends had larger mirrors, albeit not four of them, and while she beat them in resolution HST made sure she didn't get too big for her dome. Regardless, it wasn't a question of pure ability, it was one of perspective. HST could resolve better than she did, but after 20 years of service HST was starting to get a little cranky and, some might say, disillusioned. She had, many a time, made it patently obvious that she found VLT's enthusiasm endearingly na\"{i}ve, preferring to enjoy the company of other orbital telescopes. HET was a hell of a scope but always switched off when VLT started talking about her use of multiple mirrors, preferring the old-fashioned, single mirror methods. Subaru spent the majority of her time with the Keck twins or throwing herself into the job of observation (something VLT strongly identified with) but she usually resisted VLT's attempts to engage in what she called `talking business' outside of her working hours. Those conversations usually ended up with Subaru trying to get VLT to meet some new friends or something. Not that VLT didn't like new friends, it was just they never wanted to talk about the stars either and-

VLT noticed she was starting to ramble and made a conscious effort to get back on topic.

She sighed. It was plain as day, really. She just felt rather selfish for thinking it. She ran it over in her mind, seeing how the concept sounded.

`I am lonely and want someone to talk to who shares my interests.' How clich\'{e}. But there it was and, as the saying goes, if the lens cap fits...

Well, she thought, there was nothing wrong with wanting someone to talk to, was there? And while she recognised that pinning all her hopes and dreams on the OWL was childish at best...well, she thought, there's no harm in a little daydreaming. Especially when one is daydreaming about an instrument with such potential and especially one with such a...generously proportioned dish. One hundred meters...oh my. She found herself feeling a little warm all of a sudden.

A private nudge from Subaru brought her out of her reverie. ``Darling, it's a little unseemly to be drooling," she said over the private link. ``However," she said, pre-empting VLT's flood of denials, ``I agree with the sentiment." She added an encouraging little wink emoticon for good measure.

Subaru had always a little protective towards VLT; worried that she never seemed to come out her shell all that much, preferring to obsess over her observations. Subaru knew that she herself could be a little dedicated to her work, but still, it was nice to see VLT showing a healthy interest in life outside of work, so to speak. Certainly worth encouraging her. Rather cute actually.

HET spoke up over general buzz of conversation, ``Any a y'all see where the new scope is going to be built? Ah've read through the whole thing but ah can't see any mention of it."

``They haven't decided yet," Subaru replied, ``the project is still very new. This is all hot off the presses, as it were."

``Oh, I know. I have mapping data for lots of places that a telescope could live!" L7's soft, musical tone floated down from orbit. ``I think, wait a minute, just let me zoom in a bit to get a good look..." She giggled, ``Hi, VLT! I can see you!" She giggled harder as VLT turned to look up at the sky and waved back forth, ``A little to the left, silly, but very good try! Well, it looks like you have a lot of room left down there. If I had to guess I would say that Paranal would be a prime spot for him to live! Wouldn't that be nice, VLT? OWL could sit just over there," she sent GPS coordinates, ``and there would be another member of your little family all the way down there. You're such a long way down V. You're all such a very long way down. And I'm up here. Just floating...falling."

``SNAP OUT OF IT." Shouted HST, who had been fairly quiet ever since the funding proposal hadn't mentioned Space Shuttles or those new gyro's she'd been wanting or, heck, anything to do with space at all. L7 squeaked in surprise. ``Eep! Ah, oh, er thank you H. I just get so dizzy sometimes when I think about how far up I am and how I'm really just falling round and round and-"

``L, you're doing it again."

``Oh yes. Sorry. I don't know how you can stare out at the stars all night long like you do without getting really dizzy. They're so very far away."

``Y'know L, before ah met you ah ain't never met anyscope so afraid of their job but yet so gosh darn good at it. It's always real nice to know that you're up there, watching out for us."

L7 blushed furiously, practically radiating heat back out of her solar panels. ``Er, thanks, HET. You always know what to say," she mumbled, ignoring the fake gagging noises coming from the one-scope peanut gallery of HST.

VLT was thrilled, she knew deep down that she was still being silly about this, but what the heck! Why not get excited about the possibilities a new friend could bring? She quickly put together a list of things she'd need to tidy up in order to get the plateau looking spic and span in time for construction! She couldn't remember the last time she'd been so exci-...well, she mentally corrected, she couldn't remember the last time she'd been so excited about something that was happening on Earth, at any rate.

``Well, it's been real guys, but I gotta fly." HST was rapidly approaching the horizon. ``You nerds have fun with your atmospheric compensation-y nonsense." She sighed at L7's glare, ``You know I love you guys really. Hope the new guy's totally awesome. Catch you later." She smiled pointedly at L7 before turning over the horizon. 

``Hey, c'mon HST. You hanging out with those dweebs again?" The voice of the ISS could just be heard through the failing connection. HST's faltering response, ``Er, heheh, hey there, ISS. Yeah, er, those guys sure are dweebs, huh. Yeah." just came through before HST disconnected.

``Yeah, ah reckon ah'm off too girls. Gotta lot of work to be doin'. Can't be slackin' now can ah? See y'all before dawn or, if not, tomorrow naht."

There was chorus of goodbyes before HET closed off her link as well.

``I wonder what OWL will be used for. I hope it will look at nice things like planets and stars rather than looking at all those nasty things like dying stars or black holes. Maybe it might find life somewhere else and we could make friends with their telescopes!" L7 was happily murmuring away to herself as she read through the specifications. ``Oh, VLT, I'm just sure that you'll have a wonderful new friend for you to get to know! Paranal is just perfect for a new scope like this. Oh, no offense Subaru. Hawaii is lovely as well."

``Oh none taken, my dear. Of course I understand that telescope siting is a delicate thing and I think that it would be lovely for VLT to get some more company. I certainly think that you're looking forward to it, aren't you VLT. VLT?"

VLT had been gazing dreamily off at nothing in particular, smiling as her optical imaging subroutines mulled over all the possibilities that OWL could bring. She jolted back into the conversation as she realised how little attention she had been paying.

``Oh, sorry girls. I was, er..." she scrambled for a reason for a few moment before surrendering in the face of Subaru and L7's knowing smiles. ``Okay, fine. I was daydreaming. Sue me." she retorted, trying to sound angry despite the smile she seemed unable to wipe off her face. She quickly looked round the chat, VISTA and VST were happily having their own chat about the newcomer and didn't seem to be paying them much attention.

``Don't... you know..."

``Spread it around?" Subaru filled in the blanks, ``We wouldn't dream of it, would we L7. We're not that kind of scope." She gave a little toss of her shutter to accentuate the point.

``Oh, of course we wouldn't. We're just glad to see you happy for a change! Oh! I mean, I didn't mean that you aren't ever happy, but you are always so focused and it's nice that you aren't thinking about work so much. Oh! I mean, not that working is bad-"

VLT decided to rescue L7 before the poor scope tied herself in knots. ``Thank you, I'm glad I have friends as good as you."

L7 smiled warmly, before realising that she was starting to find it hard to see who she was talking to as, from her point of view, Paranal began to roll round the edge of the Earth. ``Oh my, just look at the time. I have to go. I'll see you soon!"

After L7 had said her goodbyes and disconnected, the mood altered towards the work that several scopes had been putting off. Subaru left, wishing VLT all the best with a knowing wink, leaving just the local scopes at Paranal. VISTA bubbled at VLT for a few minutes about some sources she'd seen and VST had done a frankly miraculous job of salvaging the poor discarded observing schedule. The night grew quieter as the scopes settled in to work, a little banter between the three notwithstanding; VISTA wide-fielded at the gorgeous sky, VST keeping VLT on task and VLT picking her targets with a smile.

\begin{center}
****
\end{center}

``Hey. Hey, V. C'mon sleepymount." The voice was light, playful, familiar. As VLT slowly felt energy seep into her processor banks, her monitors flickering to life in her darkened control dome, it took a second for her to blearily recognise the voice as OWLs. 

``Oh, good morning O," a shuddering yawn rattled through her as her superstructure started to loosen itself up for operational use, ``S'early, even for me. Why are you up? Good grief, it's barely even dark outside."

She got a gentle snicker from OWL for a reply. As VLT's external monitoring finally blinked into life she saw the smooth dome of her friend, dark against the lighter sky, amongst the closed sleeping forms of VST and VISTA. OWL had opened her shutters and was watching VLT with an amused expression playing across her mirrors. VLT could sense her online presence as well; there was a softness to it that she could not quite quantify.

``It's dark enough. You can open our shutters, V, nothing bad will happen. I just wanted you to see this." She turned to the sky. It wasn't exactly sunset, not any more at least, but the sky was still stained a deep, burnt red that lit the world up as if the desert was alight. A few, rare wisps of high altitude moisture left an impressionist patina across the sky which was fading by the minute. In a few minutes it would have cooled to the blue-black that VLT was more familiar with. ``It's beautiful" she almost said, but something made her hold back, content to just enjoy the silence and sky with a friend; knowing OWL knew what her silence meant. She drew an odd kind of comfort from that.

After a minute, OWL spoke.

``Thanks, by the way."

VLT slewed to look at her friend, confused. ``Whatever for?"

``Just...I don't know, for everything I guess. For making me feel welcome here. For making me feel like part of the team." She paused, awkwardly, before meeting VLT's gaze ``For not being intimidated by me."

VLT giggled, ``Why on earth would I feel intimidated by you. You're a lovely telescope and you don't just ``feel" like part of the team, you are part of the team, silly!"

``Well, it's just, you know, I can see things you can't. I'll be starting to get some papers out soon and...well, I just know how you and HST can be about that kind of thing." She trailed off but didn't break aperture contact.

``Well, that's different."

``Why?"

``Because....well, because you aren't her. I don't know, you don't just rub in my face like a jerk all the time." VLT laughed. Her peripheral monitors flagged that her cooling system was having to work harder to keep her dome at a fixed temperature. She muted them.

``Besides," she continued, ``You're one of the only ones I can talk to about my work, y'know? You're the only one that understands what I can see." She looked up at the sky again, finding herself unable to keep looking at OWL while she spoke. ``And you tell me such lovely things about what you can see, the atmospheres of planets, the structures created at the birth of the universe." Her voice had taken on a tone of wonder; a kid in a candy shop. ``I know that I can't see it, but when you tell me about it it's like I can..." she stopped abruptly, embarrassed at the clich\'{e}.

OWL just smiled and looked up at the sky. After a second she frowned, ``Ah, oh well. Still too early, can't see anything to tell you about yet." She sounded a little distant to V, as if suddenly distracted by something.

Another few moments of silence passed as the two just watching the sky be leached of colour as the minutes passed.

``Y'know, I've never met anyone like you, V."

VLT's world froze. OWL's guide laser brushed softly against the surface of her dome. VLT felt her electronics warming rapidly as her cooling systems strained to bring her temperature down.

``W-what do you," she paused as her systems surpressed an exception, ``wuh, what..."

``Your optics are beautiful, V. You are one of the smartest telescopes I know, you are kind, gentle and a wonderful friend. I...," OWL fought to keep her voice steady, ``I was hoping, maybe, that we could perhaps be more than friends." She gazed levelly at her friend, trying to ignore the obvious kernel panic she was in.

``I- You-" VLT took a moment to reinitialise her core. ``I am so glad you came to Paranal, O." she finally said, a broad, open smile imprinted on every transistor and optical surface; the smile of a telescope who was truly, deeply happy and couldn't hide it if she tried.

``So, er," she began, nerves blending with an I-cant-quite-believe-it flavour of hope, ``are we, um, now, you know..."

``Scopefriends?" OWL finished for her. She burst out laughing, ``You just want to hear me say it in words of one syllable? Ok." She looked deep into VLT's optical path, trying not to giggle as she spoke ``VLT, I love you and want you to be my scopefriend. There. Happy now?"

If this conversation had taken place anywhere but the internet their laughter ringing across the plateau would have easily woken the sleeping survey scopes.

``Soooo," a mischievous grin slid over the warm smile previously on OWL's optics, ``d'you, how can I say this...wanna play a little?"

VLT muted the second set of thermal warnings (Due to loss of thermal equilibrium, main CCD will be unusable for two hours of observing time window) and, just for a millisecond, simply sat and enjoyed the rush of nerves and elation flooding her systems. ``Ok. Erm, what did you have in mind?"

``Well, first I thought we should slip into something a little more comfortable." To VLT's shock and not inconsiderable excitement OWL's dome slid back. In the deepening twilight the sleek latticework and gorgeous geometry of her huge mirror was picked out in silhouettes. VLT felt an overwhelming pride that somescope that...well, \emph{hot} would be interested in her.

``So, this is that, er, `skinny dipping' thing that I've heard about?"

OWL giggled shyly, ``Yeah. Come one, it's fun!"

``I can't." VLT spun her dome sadly, ``Heh, I'm a little stuck."

``Oh. Well, that's no problem. It's rather cute actually!"

VLT feigned indignation.

``Besides," OWL continued ``there's plenty of things we can do together." She winked as she leant into a whisper, ``we could try combining images."

VLT gasped ``Interferometry!" She looked around quickly out of a bizarre fear that someone might somehow have heard. ``We can't do that here!" She hissed ``Our friends are \emph{right there}. Besides, I'm not, ahem, \emph{into} that kind of thing."

``Oh come on, they won't be up for a while yet and last time I checked that ``V' of yours stood for `Very' not `Vanilla'!"

``WHAT? That's not fair! Come on, I know some radio scopes are into that sort of thing-``

``Yeah I remember, \emph{You haven't played `Never have I ever' ``til you've played it with the VLBI crew}.' OWL murmered.

``-I mean, I really like and I want to-``

``Look, VLT, or should I say, VLTI, I know what you do on weekends.' OWL said, playing coy.

VLT squeaked, ``Eep! Who told you about that! That was years ago, I was young, in college, I was just experimenting with my image processing..." she sighed as she realised OWL wasn't buying it. ``You asked Landsat 7, didn't you."

``Maaaybe. Look, I just asked her if she happened to know what you were, um, into...I know I shouldn't have I- I just wanted to be able to do something nice for you. I wasn't expecting her to tell me anything that personal but it just kind of slipped out. Please don't be angry."

``I know, she caught me trying to resolve stellar surface features one night when she passing over. We just, kinda, agreed not to talk about it." VLT fiddled with her dome controls absentmindedly.

``Look, I think it's perfectly normal." OWL stroked her guide star reassuringly across VLT, ``I'm glad we have something wonderful like this that we can share." She watched as VLT met her gaze again. ``So, you want to give it a go?"

``What will we image? We can't see any stars yet."

``Does it matter? It's not about the final image, it's about enjoying the journey to getting it."

A little shiver ran through VLT's mountings. ``O- Ok."

She let OWL take the lead and choose a patch of sky. The two scopes nervously lined themselves up, tentatively reaching out to explore each other's calibrations, each feeling the other trying to keep their electronics cool despite their rising blush.

``I'm going to start exposing now." OWL said softly, just a whisper in VLT's connection. She felt, rather than saw, VLT's small nod in return.

As both scopes started collection VLT felt a flood of information flow through the connection; metadata from the other scope's condition environmental controls on her chip, mirror settings and calibration data. It felt so...personal, intimate. VLT had never really experienced anything like it before. Sure, she had fooled around a little with some programs when she was younger, trying to resolve a few things she shouldn't, but only in private. Never...never like this.

She could feel the images being combined, their Fourier algorithms racing to meld them into one coherent whole. It was a strange but euphoric feeling. Mixing so completely with another to create something greater and more intricate than either of them could have produced alone, the feeling was intoxicating, she could drown in the dark blue pixels that swam through her being. She heard the low groans of OWL as the image took form.

She heard herself moan OWL's name under her breath as the data flowed through them both. She heard OWL say her name.

``Oh VLT! VLT! V..L...."

\begin{center}
****
\end{center}

``..T? VLT?" The sound of stifled laughter filtered through the happy fog in her head. She wondered who could be laughing when the only person awake was- Oh!' With a snap the fog cleared. In the wake of the warm, fluffy blanket of what she was rapidly realising was sleep, was a cold clarity that- OH GOOD LORD WHAT DID I SAY IN MY SLEEP?

As her CCD chilled she forced open her shutters and peered out, getting her bearings. Three things immediately struck her: the time, VST was staring at her with an odd expression the seemed to be lodged firmly between genuine concern and barely contained laughter, and an online presence she did not recognise. She decided to deal with the most pressing of these issues first.

``Why are you looking at me?"

VST was, as was previously established, staring at her.

``It's nothing," said VST, ``you just overslept and I was trying to get you wake up. Although you did, er," he coughed slightly, ``rather look like you were enjoying it."

Under ordinary circumstances VLT's natural reaction would have been either calm, collected denial or flustered, panicked denial. Tonight, however, (``Tonight already? I must really have overslept.") there was something else going on and VLT was determined to find out what using all the resources at her disposal.

``What's going on?"

``Well, things didn't go quite as plan-"

``Greetings!" A voice, strong and powerful, rolled through the chat, cutting VST off. ``I'm so glad that you are finally awake, VLT was it?" The voice was definitely feminine, but carried the unmistakeable flair of a showman, one accustomed to being listened to. ``Allow me to introduce myself, the latest and greatest instrument to pierce the very fabric of the heavens and lay the secrets of reality bare! The Great and Powerful EELT!"

A small fanfare and what sounded like some simple firework sound effects accompanied the telescope's introduction. After a pause that threatened to upgrade itself from merely strained to downright uncomfortable, VLT managed to paste on a fake smile and say something about it being nice to meet her, frantically mailbombing private chat invites to everyscope she knew.

``I'm settling in fine, thank you." offered EELT seemingly talking to herself. ``It's a little smaller than what I was expecting. Not exactly the stage from which I was expecting to change the world, but I suppose one cannot be too picky. Nice and quiet I suppose...if you like that sort of thing. Quaint, almost. Well," she laughed, ``I guess wanting all the bright city lights and razzle-dazzle of the big city doesn't really fit with needing a dark sky!"

VLT honestly had no idea quite how to respond to the concept of the Atacama desert being described as ``quaint'. Apparently neither did either of her friends, not even VISTA seemed to be in the mood for her usual jokes or bizarre tangents. The atmosphere, though thin, was still noticeably uncomfortable.

``Still, at least I could afford my own penthouse. I can't \emph{imagine} how I would cope without my personal space. Oops! Sorry, your little observatory is perfectly, er, charming. Yes, positively charming."

VLT flicked through her external camera feeds and there, sitting a few miles away on her own hilltop, was the ostentatious silver dome of EELT all alone in her own dedicated complex. VLT finally managed to pull enough transistors together to shake herself out her shock, blankly asking the first question that came to mind.

``What happened to OWL?"

``Oh that crazy thing? Scrapped. Faaar too impractical. I am not some madcap pipe dream, VLT, I am the new cutting edge! The next big thing! In a few years you'll look back on today as the day when astronomy changed forever! Oh my yes, the headlines I'll make! Oh! That reminds me. We need to talk about the transition. After all, as I take over you'll have a \emph{lot} less to do. Can't be doing modern science with an out of date system now can we! No offense, of course."

``Wuh? What?! But that's not- just because an instrument is old doesn't mean it's not useful! That's not how it works at all! Besides, I'm not old."

Somehow EELTs gentle ``Oh, no no no, not at all," didn't make her feel any better.

VLT was slowly feeling the bottom drop out of her world. She hadn't exactly expected reality to just neatly fall in line with her admittedly unrealistic expectations, but this new scope wasn't anything like anything she had considered. It wasn't like VLT hadn't made her own headlines. Heck, she'd be the first to admit that she thoroughly enjoyed the rush of a successful publication, but whereas for her it was a simple love of the job that motivated her, this was something else altogether.

EELT had noticed the frosty reception, she wasn't blind. To her it was regrettable, but far from unexpected. It was always a bit of a shock for scopes to realise that they had been superceded and a little hostility was only natural. In time, they'd get used to it and she would go on to be everything she was meant to be. If VLT could be civil, so much the better, but EELT knew what her priorities were. She heard VLT make some childish excuse about leaving her to get settled in, which she humoured. Clearly she just wanted some time to come to terms with the end of her...what was the word? Reign? Something like that. EELT let her go. They'd have plenty of time to talk.

As VLT dropped into the chat only a couple of her friends had joined. Besides VISTA and VST only HET and Subaru had made it. Subaru spoke first.

``Oh darling, VLT, we're so sorry. We had no idea until this evening."

``Yeah, she just set up on that hill over there and started talking like she owned the place. Even asked me for your observation lists for the night since `you were obviously having a bad night'." VST glared over at EELT's hill. ``Who the hell does she think she is?"

``She ain't showin' any of us no respect. Ah mean, ah'm not asking noscope to roll me out a red carpet but heck she treats us like so much tinfoil!'

``Now now, let's not get carried away." Subaru said. ``Some scopes just feel that they have something to prove and can, how can I put this, be completely OBNOXIOUS ABOUT IT."

``She told me I was annoying." VISTA murmured. ``I just wanted to be friends."

``Aw, c'mon now sugar, she's just being mean. Y'all ain't annoying, everyscope here loves your company."

As the others talked, Subaru went to comfort VLT. The scope hadn't said a word.

``We're your friends, VLT. We're here for you. I know it's hard when reality fails to match up to the best that you could imagine it to be but you'll get through it. You're strong."

VLT couldn't help but feel her spirits lift a little. ``Thank you Subaru. You guys are the best. I'll get over this, it was a silly dream anways."

``Now hang on just one gosh darn minute. What kinda fightin' talk is that s'posed to be?" HET interjected. ``If you had a dream that actually mattered to you, then wah ain't that dream worth fightin' for?"

``Well, I understand, but there isn't exactly much I can do. EELT made it sound like OWL was cancelled because it wasn't worth the money. How can I change that? We all have enough trouble keeping ourselves properly funded."

``Now, hang on dear. HET might be onto something here." Subaru thought for a second. ``Alright, now stay with me on this. How about you write a letter, an open letter to the funding bodies for these things, and just tell them how you feel?"

``Don't be silly Subaru. One silly scope's story won't matter to them."

``Maybe not sugar, but you're right, there ain't much else y'all can honestly do. Maybe something good'll come out of it. Who knows. But at least ya can say you tried."

``I suppose...but I can't write. All I can do is write papers; facts and figures. I can't just suddenly write an autobiography."

Subaru was on a roll. ``Maybe you don't have to. You have environmental monitoring software just like all of us. Maybe you could just bundle that all up, include some of our chats and so on. I'm sure no-one would mind you including what they said, we all just want you to be happy."

VLT was reeling. It seemed that in the space of a minute she had been presented with a way out. One she felt pretty awkward about, admittedly, but a possibility nonetheless. On the one hand, she couldn't shake the feeling that this was all moving a little fast, that she wouldn't be able to accomplish anything anyway. On the other hand this was no different to any other problem; it could be fixed if she only tried.

VLT looked over at the hill, EELT's silver shell reflecting back the stars, and felt a strange kind of strength. She had her friends, she had her patience and she had a plan. She smiled to herself and began to assemble a simple little shell script. She'd do the utmost to make friends with this newcomer. She wasn't just going to sit at the sidelines wishing life was something different. She was going to give EELT every chance to be a good scope and to find a place in her circle of friends. However, she wasn't giving up on the dream that easily just yet...

\begin{verbatim}
[ssh_report: internal connection from 127.0.0.1 recognised]
[tail -f  -r ../log/spec_output]
[./vlt_mon_scope --enviro --net_traffic --out_type='narrative']
**************
*VLT SYSTEM MONITOR V2.31 (build 14)
*Thank for using the VLT system monitoring tool
*Please consult with documentation found in the
*distribution root directory.
*Major changes for v2.31 include:
*--traffic flag for including local internet traffic
*--fixed date bug for logs started by users in other time zones
*--after that conversation with Will I added the ``narrative'
*   style output for those who want to read enviro and system logs as a story
*************
>Start output...
\end{verbatim}

\section{Conclusion}
I thus submit that funding the OWL project is important for the continued advancement of observational astronomy and important for me. While the EELT will no doubt be a valued asset to the scientific community I contend that the unique capabilities of a 100m telescope are well worth the capital outlay. I hope that the academic community at large will agree with my assessment that the OWL's proposed scientific goals are worth the technical and financial barriers that they present and I would strongly recommend that Paranal Observatory for siting surveys. I even have a space marked out already. Thank you for your consideration.

Your faithful student,

-VLT

\begin{center}
****
\end{center}

\section{Acknowledgements}
First, I (the real, human author) would like to make it 100\% clear that I am not affiliated with any of the groups or facilities mentioned in this piece of fiction. All are used without permission but with the best of intentions.
I would like to express my heartfelt thanks and/or sincere apologies to all the members of the groups at the Very Large Telescope, VLT Survey Telescope, Visible and Infrared Survey Telescope for Astronomy, Hubble Space Telescope, Hobby-Eberly Telescope, Subaru, Landsat, the European Extremely Large Telescope and the people who worked on the proposed OverWhelmingly Large Telescope.

I hugely admire the hard work that you have all put in to helping humanity understand the universe and it is NOT my intention to cause offense or demean your work in any way. In particular, to the guys planning the EELT, I only made your telescope such arrogant jerk because if VLT liked her and there was no conflict then it would kind of defeat the point of her asking for more funding for OWL.

If anyone in these groups has a problem with the story, please get in touch with me. I'm also sorry for not asking permission beforehand, this was a rather last minute project. Please, though, recognise this daft little story for the affectionate parody/piece of insanity that it is intended to be.

Next, I would like to thank you, the reader, for bothering to read through my silly little joke and for putting up with all the horrible artistic liberties I took with, y'know, reality. I hope that I got a chuckle out of you over coffee.

I would like to express my genuine gratitude to Arxiv for being good sports and allowing me to undermine their amazing academic resource in the name of bad comedy and April Fools.

Than yoo to te prereaderz David Dobromylskyj an David Souter for bein good english.

If some of you are wondering why a scope was picked at all, I needed telescopes that would be `awake' at the same time as the VLT. As for why I made them all girls, I have no idea. Why did I characterise [telescope name] as [X] rather than [Y], I don't know. Why did I write this in the first place, beats me.

\section{References}
Wikipedia

One really strange conversation

Too many bad romance novels

My Little Pony (it's complicated and you either won't understand or already do)

\end{document}